\documentclass[final,3p,times,sort&compress]{elsarticle}

\usepackage{multirow,setspace,amssymb,amsmath,graphicx,color,rotating,subfigure,url,lineno,natbib}
\journal{Physica A} %%  change this journal name and put the correct one

\begin{document}

\begin{frontmatter}

\title{Complex stock trading network among investors}
\author[BS,SS,RCE]{Zhi-Qiang Jiang}
\author[BS,SS,RCE,RCFE]{Wei-Xing Zhou \corref{cor}}
\ead{wxzhou@ecust.edu.cn}
\cortext[cor]{Corresponding author. Address: 130 Meilong Road, P.O.
Box 114, School of Business, East China University of Science and
Technology, Shanghai 200237, China, Phone: +86 21 64253634, Fax: +86
21 64253152.}
\address[BS]{School of Business, East China University of Science and Technology, Shanghai 200237, China}
\address[SS]{School of Science, East China University of Science and Technology, Shanghai 200237, China}
\address[RCE]{Research Center for Econophysics, East China University of Science and Technology, Shanghai 200237, China}
\address[RCFE]{Research Center on Fictitious Economics \& Data Science, Chinese Academy of Sciences, Beijing 100080, China}

\begin{abstract}
We provide an empirical investigation aimed at uncovering the statistical properties of intricate stock trading networks based on the order flow data of a highly liquid stock (Shenzhen Development Bank) listed on Shenzhen Stock Exchange during the whole year of 2003. By reconstructing the limit order book, we can extract detailed information of each executed order for each trading day and demonstrate that the trade size distributions for different trading days exhibit power-law tails and that most of the estimated power-law exponents are well within the L{\'e}vy stable regime. Based on the records of order matching among investors, we can construct a stock trading network for each trading day, in which the investors are mapped into nodes and each transaction is translated as a direct edge from the seller to the buyer with the trade size as its weight. We find that all the trading networks comprise a giant component and have power-law degree distributions and disassortative architectures. In particular, the degrees are correlated with order sizes by a power-law function. By regarding the size executed order as its fitness, the fitness model can reproduce the empirical power-law degree distribution.
\end{abstract}


%      {}{Economics; econophysics, financial markets, business and management}   \and
%      {}{Networks and genealogical trees} \and
%      {}{Systems obeying scaling laws}

\end{frontmatter}

\section{Introduction}
\label{Sec:intro}

Most people believe that the global economic and financial systems exhibit more and more remarkably intertwined nature with a continuing increase in economy globalization using different measures \cite{Li-Jin-Chen-2003-PA,Garlaschelli-DiMatteo-Aste-Caldarelli-Loffredo-2007-EPJB,Ausloos-Lambiotte-2007-PA,Miskiewicz-Ausloos-2008-PA,Song-Jiang-Zhou-2009-PA,Pukthuanthong-Roll-2009-JFE,Schiavo-Reyes-Fagiolo-2010-QF},
although some measures give a different scenario \cite{Pukthuanthong-Roll-2009-JFE,Miskiewicz-Ausloos-2010-PA}, which provides a diffusion path for the US subprime mortgage lending crisis triggering the current global financial and economic crisis, mainly in the developed economies \cite{Schiavo-Reyes-Fagiolo-2010-QF}. There is no doubt that studying the structure and dynamics of financial and economic networks will provide revolutionary insights to our understanding of the evolution of financial and economic systems, which have predictive implication for policy makers \cite{Schweitzer-Fagiolo-Sornette-VegaRedondo-Vespignani-White-2009-Science}. The ideal picture would be that we have a unified network for all economic units and the detailed information of their interactions. Unfortunately, such a database is almost impossible to build. Nevertheless, the importance of network analysis of financial and economic systems is self-evident, especially in the Econophysics community.

In recent years, complex network theory has witnessed a flourishing progress in many interdisciplinary natural and social fields \cite{Watts-Strogatz-1998-Nature,Barabasi-Albert-1999-Science,Albert-Barabasi-2002-RMP,Newman-2003-SIAMR,Dorogovtsev-Mendes-2003,Boccaletti-Latora-Moreno-Chavez-Hwang-2006-PR}.
Quite a few network patterns differing from random graphs have been revealed, such as small world \cite{Watts-Strogatz-1998-Nature,Amaral-Scala-Barthelemy-Stanley-2000-PNAS}, scale-free degree distribution \cite{Barabasi-Albert-1999-Science}, community \cite{Ravasz-Somera-Mongru-Oltvai-Barabasi-2002-Science,Guimera-Amaral-2005-Nature,Krause-Frank-Mason-Ulanowicz-Taylor-2003-Nature,Newman-2004-PNAS}, and rich club \cite{Zhou-Mondragon-2004-IEEE,Colizza-Flammini-Serrano-Vespignani-2006-NP,Jiang-Zhou-2008-NJP}, to list a few.
Complex financial and economic networks can be classified in three categories. In the first category of networks, the nodes present financial or economic agents (economies, companies, financial institutions, traders, et al) and a link is drawn between two nodes if they have certain interactions (such as investment, trade, lending, economic cooperation, et al) \cite{Serrano-Boguna-2003-PRE,Boss-Elsinger-Summer-Thurner-2004-QF,Garlaschelli-Loffredo-2004-PRL,Garlaschelli-Battiston-2005-PA,Garlaschelli-Loffredo-2005-PA,Hochberg-Ljungqvist-Lu-2007-JF,Kogut-Urso-Wakler-2007-MS,Battiston-Rodrigues-Zeytinoglu-2007-ACS,Fagiolo-Reyes-Schiavo-2008-PA,Iori-Masi-Precup-Gabbi-Caldarelli-2008-JEDC,Fagiolo-Reyes-Schiavo-2009-PRE,Barigozzi-Fagiolo-Garlaschelli-2009-XXX}.
In the second category of networks, the nodes are agents, each of which has a time series recording its behavior and a link is drawn due to the correlation between two time series \cite{Mantegna-1999-EPJB,Bonanno-Vandewalle-Mantegna-2000-PRE,Bonanno-Lillo-Mantegna-2001-QF,Onnela-Chakraborti-Kaski-Kertesz-2002-EPJB,Onnela-Chakraborti-Kaski-Kertesz-2003-PA,Tumminello-Aste-DiMatteo-Mantegna-2005-PNAS,Tumminello-DiMatteo-Aste-Mantegna-2007-EPJB}. The third category contains networks converted from individual financial time series \cite{Yang-Yang-2008-PA,Ni-Jiang-Zhou-2009-PLA,Yang-Wang-Yang-Mang-2009-PA,Liu-Zhou-2009-XXX,Qian-Jiang-Zhou-2009-XXX}, based on different mapping methods such as phase space embedding, visibility algorithm, sub-series correlation, recurrence, and n-tuple occurrence \cite{Zhang-Small-2006-PRL,Li-Wang-2006-CSB,Li-Wang-2007-PA,Zhang-Sun-Luo-Zhang-Nakamura-Small-2008-PD,Xu-Zhang-Small-2008-PNAS,Lacasa-Luque-Ballesteros-Luque-Nuno-2008-PNAS,Marwan-Donges-Zou-Donner-Kurths-2009-PLA,Donner-Zou-Donges-Marwan-Kurths-2010-NJP}.

The complex financial and economic networks of first type range from the macroscopic level of countries to the microscopic level of traders in markets. Trading networks at the microscopic are less studied, since the detailed information of trader identities and their transactions is usually unavailable to researchers with merely a few exceptions. Kyriakopoulos et al have investigated the statistical properties of the transaction network of all major financial players (423 accounts) within Austria over one year \cite{Kyriakopoulos-Thurner-Puhr-Schmitz-2009-EPJB}. They found that the directed network is disassortative and the random matrix analysis is able to identify an account with financial misconduct. Tseng et al conducted a series of web-based prediction market experiments for 97 days with 2095 effective participants and 16936 transactions to explore the structure of transaction networks and to study the dynamics of wealth accumulation \cite{Tseng-Li-Wang-2010-EPJB}. They found that \cite{Tseng-Lin-Lin-Wang-Li-2010-PA}, the trading networks are scale-free and disassortative. These empirical properties can be well reproduced by a zero-intelligence agent-based market model, outperforming the zero-intelligence plus model \cite{Cliff-Bruten-2000-CEFE} and the Gjerstad-Dickhaut model \cite{Gjerstad-Dickhaut-1998-GED}. Recently, trading networks constructed from the September 2009 E-mini S\&P 500 futures contracts were analyzed by Adamic et al \cite{Adamic-Brunetti-Harris-Kirilenko-2010-SSRN}. They found that there are contemporaneously correlated relationship between network metrics (centralization, assortative index, clustering coefficient, and large strongly connected component) and financial variables (returns, volume, duration, and market liquidity). By constructing trading network from the real trading records from Shanghai Futures Exchange, Wang and Zhou found that the futures trading networks exhibit such features as scale-free, small-world effect, hierarchical organization, and power-law betweenness distribution \cite{Wang-Zhou-2010-XXX}.

To our knowledge, studies about the trading networks of all traders involved in the transactions of individual stocks are still lack. In this work, we attempt to fill this gap by investigating the trading networks using the order flow data of a highly liquid stock named Shenzhen Development Bank listed on the Shenzhen Stock Exchange, an emerging market. In consistence with many empirical social complex networks, the stock trading networks exhibit power-law degree distributions. This indicates that there are very few investors with many trading connections while there are many investors with only a few links. We further find the disassortative architecture of the trading networks and the power-law correlation between order sizes and degrees, which motivates us to employ the fitness model to explain the apparent power-law degree distribution.

The paper is organized as follows. In Sec. \ref{Sec:RulesData}, we briefly describe the trading rules of Shenzhen Stock Exchange and the data sets we adopt. Section \ref{Sec:Transaction} defines ``trade'' and its size and studies the empirical distribution of trade sizes. We provide an overview of the trading networks in Section \ref{Sec:Network} and investigate their global properties in Section \ref{Sec:Degree}. The fitness model is adopted to uncover the observed power-law degree distribution in Section \ref{Sec:Model}. At last, Section \ref{Sec:Conclusion} concludes.

\section{Trading rules and data sets}
\label{Sec:RulesData}

Shenzhen Stock Exchange, which constitutes the stock markets in Chinese mainland with Shanghai Stock Exchange, provides the market place and facilities for securities trading. Usually, the exchange is open for trading from Monday to Friday, except public holidays and other days as announced by the exchange. On each trading day, open call auction is held between 9:15 and 9:25, followed by cool period from 9:25 to 9:30. Before July 1, 2006, The continuous trading was from 9:30-11:30 and 13:00-15:00. Note that after July 1, 2006, the continuous trading is from 9:30-11:30 and 13:00-14:57, followed by closing call auction between 14:57-15:00.

Generally speaking, investors can submit market orders and limit orders and cancel unfilled orders. However, only limit orders are allowed for submission in Chinese stock markets in 2003. When the price of an incoming limit order is not less aggressive, transaction occurs. The incoming limit order might be partially filled or fully filled. A fulled filled order or the filled part of a partially filled order can be regarded as an effective market order, while an unfilled order or the unfilled part of a partially filled order is treated as an effective limit order. Without loss of clarity, we term them as market order and limit order for simplicity. During different trading periods, the trading system will take different treatments to the submitted orders. Our first step is to reconstruct the limit order book based on the order flow data for each trading days. Hence, it is necessary to elaborate to the order executing regulations for each trading period (opening call auction, cool period, and continuous double auction). It is worth pointing out that all the submitted orders are sorted according to the principle of price-time priority.

During the opening call auction, all the submitted orders are added into the limit order book. This means that the trading system regards all the submitted orders as limit orders. Cancelation order submitted before 9:20 will be executed instantly after its submission and cancelation orders submitted between 9:20 to 9:25 are frozen which take into effect only after the call auction. At 9:25, the call auction transactions occur based on the principle of maximizing the trading volume. The remaining orders in the order book matched with the cancelations, which are submitted between 9:20 and 9:25, are canceled by the trading system. The unexecuted orders automatically enter the cool period. During the cool period, the incoming market orders will be executed instantly by matching the orders on the opposite side, while the limit orders are added into the limit order book. However, the cancelations are not executed right after it was submitted. At 9:30 am, cancelations of remaining orders on the book will be centralized processed by the trading system. The unexecuted orders automatically enter the continuous double auction period. During the continuous double auction, the limit orders are added into the limit order book. All the market orders and order cancelations will be executed right after they are submitted. Before the closure of market, the unexecuted orders are automatically canceled by the trading system.

Our analysis is based on the order flow data for a stock named Shenzhen Development Bank Co., LTD (A-share, stock code: 000001) during the whole year of 2003 (totally 237 trading days) \cite{Gu-Chen-Zhou-2007-EPJB,Gu-Chen-Zhou-2008a-PA}. Each entry (or event) of the data contains the time of order placement or cancelation that is accurate to 0.01 second, direction of buy and sell, order size, limit price, and the encrypted identity of the trader. Table~\ref{Tb:OrderFlowData} reports a segment of taken from the order flow data recorded on 2003/01/02. The indicator is a buy-sell identifier, which identifies whether a record is a buy order, a sell order, or a cancelation. The database totally records 3,925,832 events submitted by 595,836 investors, including 1,718,156 buy orders, 1,595,961 sell orders, 598,750 cancelations and 12,965 invalid orders. Note that the institutional investors only account for a very small fraction (0.55\%) of total investors \cite{Mu-Zhou-Chen-Kertesz-2010-NJP}, which is irrelevant to the order splitting strategies of institutions that result in order packages \cite{Vaglica-Lillo-Moro-Mantegna-2008-PRE,Lillo-Moro-Vaglica-Mantegna-2008-NJP,Moro-Vicente-Moyano-Gerig-Farmer-Vaglica-Lillo-Mantegna-2009-PRE}. Using this database, we can rebuild the limit order book according to the trading rules and construct the trading networks for each trading day. The reconstructed transaction records are compared with the historical price and volume series to ensure the reliability of the obtained limit order book. It is evident that the size of our data set is much larger than the Austrian trading network \cite{Kyriakopoulos-Thurner-Puhr-Schmitz-2009-EPJB} and the experimental market \cite{Tseng-Li-Wang-2010-EPJB}.

\begin{table}[htp]
 \caption{\label{Tb:OrderFlowData} A segment of the order flow data}
 \medskip
 \centering
 \begin{tabular}{ccccr}
 \hline \hline
  time & ID & indicator & order price & order size \\
  \hline
  9342152 & $a$ & 28 & 10.20 & 300 \\
  9342174 & $b$ & 28 & 10.19 & 300 \\
  9342322 & $c$ & 23 & 10.20 & 11000 \\
  9342332 & $d$ & 25 & 10.22 & 4000 \\
  9342371 & $e$ & 26 & 10.21 & 300 \\
  9342382 & $f$ & 28 & 10.05 & 1000 \\
  \hline \hline
 \end{tabular}
\end{table}

\begin{figure}[htb]
\centering
\includegraphics[width=8cm]{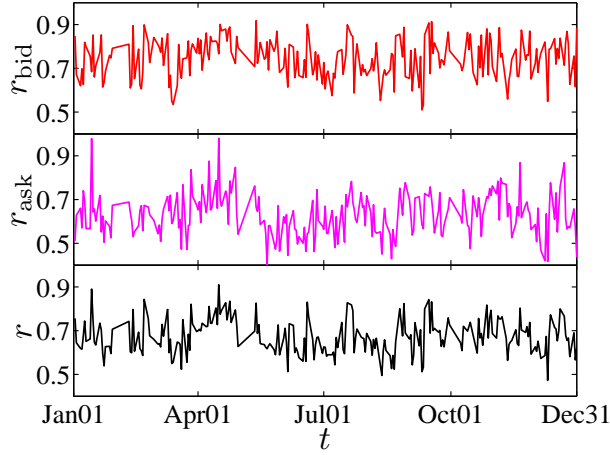}
\caption{\label{Fig:TransactionRatio} (Color online) Plots of the transaction ratios $r$, $r_{\rm{ask}}$, and $r_{\rm{bid}}$ with respect to the trading days.}
\end{figure}

We define transaction ratio $r$ to characterize how many orders out of the total submitted orders are eventually executed. The transaction ratio $r$ (respectively, $r_{\rm{ask}}$ and $r_{\rm{bid}}$) is nothing but the ratio between the number of executed orders (ask orders and bid orders) and the number of total placed orders (ask orders and bid orders). Fig.~\ref{Fig:TransactionRatio} illustrates the evolution of ratios ($r$ in the low panels, $r_{\rm{ask}}$ in the middle panels, and $r_{\rm{bid}}$ in the up panels) with respect to trading days. We find that there is no remarkably trend in the three curves and most of the values are centered around their means.  On average, 67.3\% (respectively, 68.8\%, 74.3\%) of the total orders (ask orders, bid orders) are matched.

\section{Definition of trade size and its probability distribution}
\label{Sec:Transaction}

\begin{figure}[htb]
\centering
\includegraphics[width=7cm]{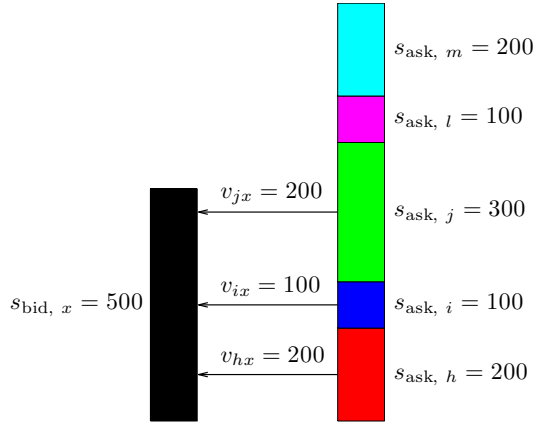}
\caption{\label{Fig:TradeSize} (Color online) Diagram illustrating the definition of trade size $v$. Five investors $h,i,j,l$ and $m$ submitted effective limit orders with sizes $s_{{\rm{ask}},h} = 200$, $s_{{\rm{ask}},i} = 100$, $s_{{\rm{ask}},j} = 300$, $s_{{\rm{ask}},l} = 100$, and $s_{{\rm{ask}},m} = 200$ at successive times $t_h<t_i<t_j<t_l<t_m$, stored as a queue at the best ask price in the ask order book. Investor $x$ then submitted an effective buy market order of size $s_{{\rm{bid}},x} = 500$ at time $t_x>t_m$. The matching of the bid order and three different ask orders leads to three different transactions.}
\end{figure}

According to the trading rules, one transaction occurs when a bid order and an ask order are matched. We define the corresponding size of order matching as trade size $v$. The trade size here is the minimum quantity of trading volume at the microstructure level of stock market and has the physical meaning of transferring a certain amount of shares from a seller to a buyer. In order to have a better understanding of the definition of trade size, we present a schematic diagram of order matching in Fig.~\ref{Fig:TradeSize}. The diagram describes the matching situation that a market bid order matches with the limit orders at the best ask price, which gives the best ask price as a transaction price. The right bar is a queue of effective sell limit orders submitted by five investors $h,i,j,l$ and $m$ at successive times $t_h<t_i<t_j<t_l<t_m$ with sizes $s_{{\rm{ask}},h} = 200$, $s_{{\rm{ask}},i} = 100$, $s_{{\rm{ask}},j} = 300$, $s_{{\rm{ask}},l} = 100$, and $s_{{\rm{ask}},m} = 200$. When an effective buy market order (represented by the left bar) with size $s_{{\rm{bid}},x} = 500$ is submitted by investor $x$, it will match three different ask orders placed by investor $h$, $i$, and $j$, which leads to the bilateral transactions between investor $h$ and $x$ with trade size as $v_{hx} = 200$, between $i$ and $x$ with $v_{ix} = 100$, and between $j$ and $x$ with $v_{jx} = 200$. We treat these as three trades, which is different from the conventional meaning of trade with size $\omega$ defined according to the aggressive side \cite{Lillo-Farmer-Mantegna-2003-Nature,Lim-Coggins-2005-QF,Zhou-2007-XXX,Mu-Chen-Kertesz-Zhou-2009-EPJB},  and we have  $\omega_x = v_{hx}+v_{ix}+v_{jx}$ in the present case.

The statistical properties of trade size $\omega$ and trading volume have attracted considerable interests in the Econophysics community, attributing to the fact that the understanding of their behaviors is particularly relevant to the price formation dynamics. The distributions of trade size $\omega$ for stock markets are found to exhibit power-law tail distributions. However, there has been a heated controversy about whether the estimated power-law tail exponents locate in the L{\'e}vy regime ($\gamma_v < 2$). By adopting different estimating methodologies, some groups find that the tail exponents are in the L{\'e}vy regime \cite{Gopikrishnan-Plerou-Gabaix-Stanley-2000-PRE,Plerou-Gopikrishnan-Gabaix-Amaral-Stanley-2001-QF,Plerou-Gopikrishnan-Gabaix-Stanley-2004-QF,Maslov-Mills-2001-PA,Plerou-Stanley-2007-PRE}, while some others disagree \cite{Eisler-Kertesz-2006-EPJB,Eisler-Kertesz-2007-PA,Racz-Eisler-Kertesz-2009-PRE,Lee-Lee-2007-PA,Zhou-2007-XXX,Mu-Chen-Kertesz-Zhou-2009-EPJB}.
Here we contribute to this literature by investigating the distribution of trade size $v$ defined in Fig.~\ref{Fig:TradeSize}.

\begin{figure}[htb]
\centering
\includegraphics[width=5cm]{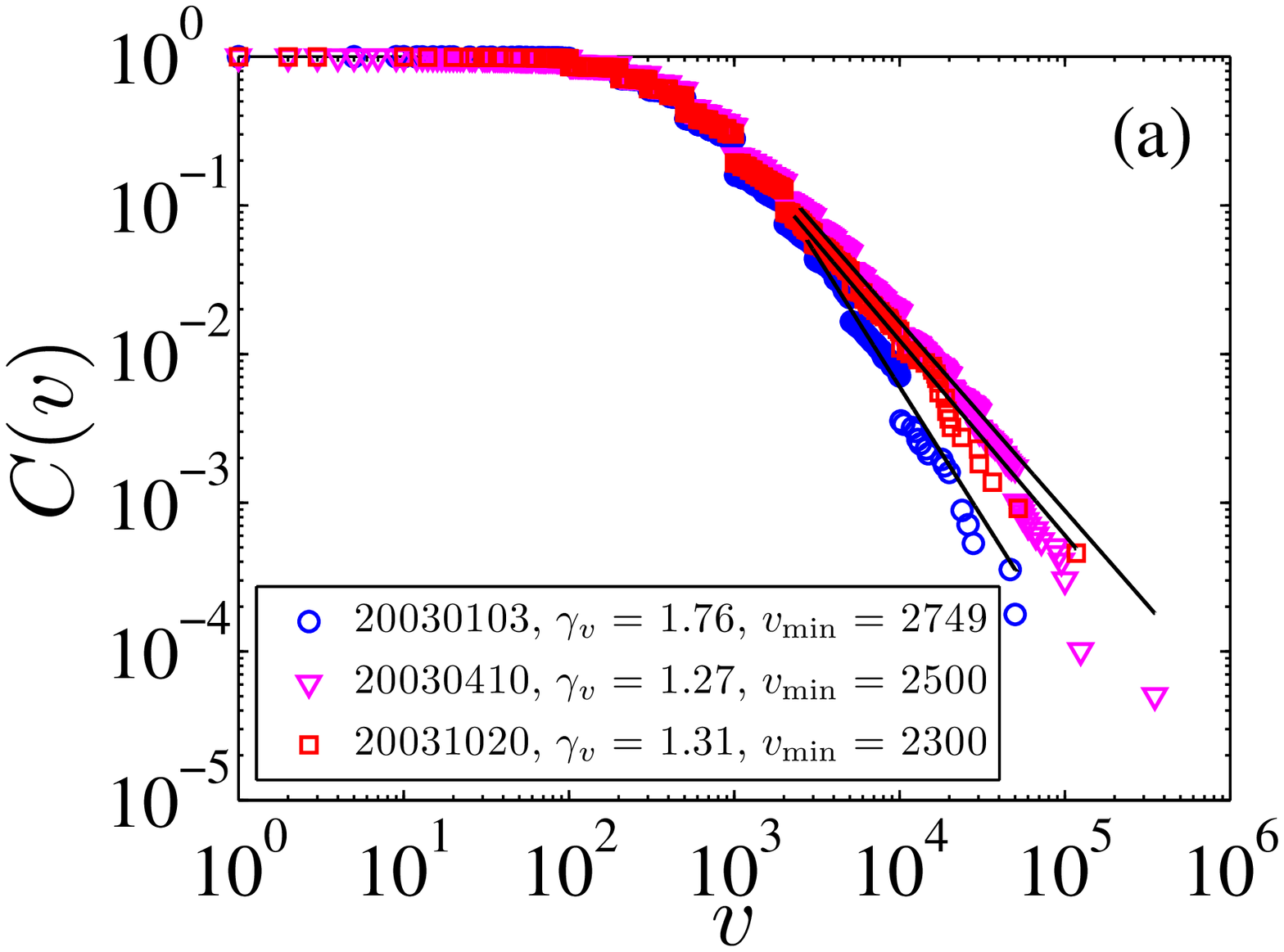}
\includegraphics[width=5cm]{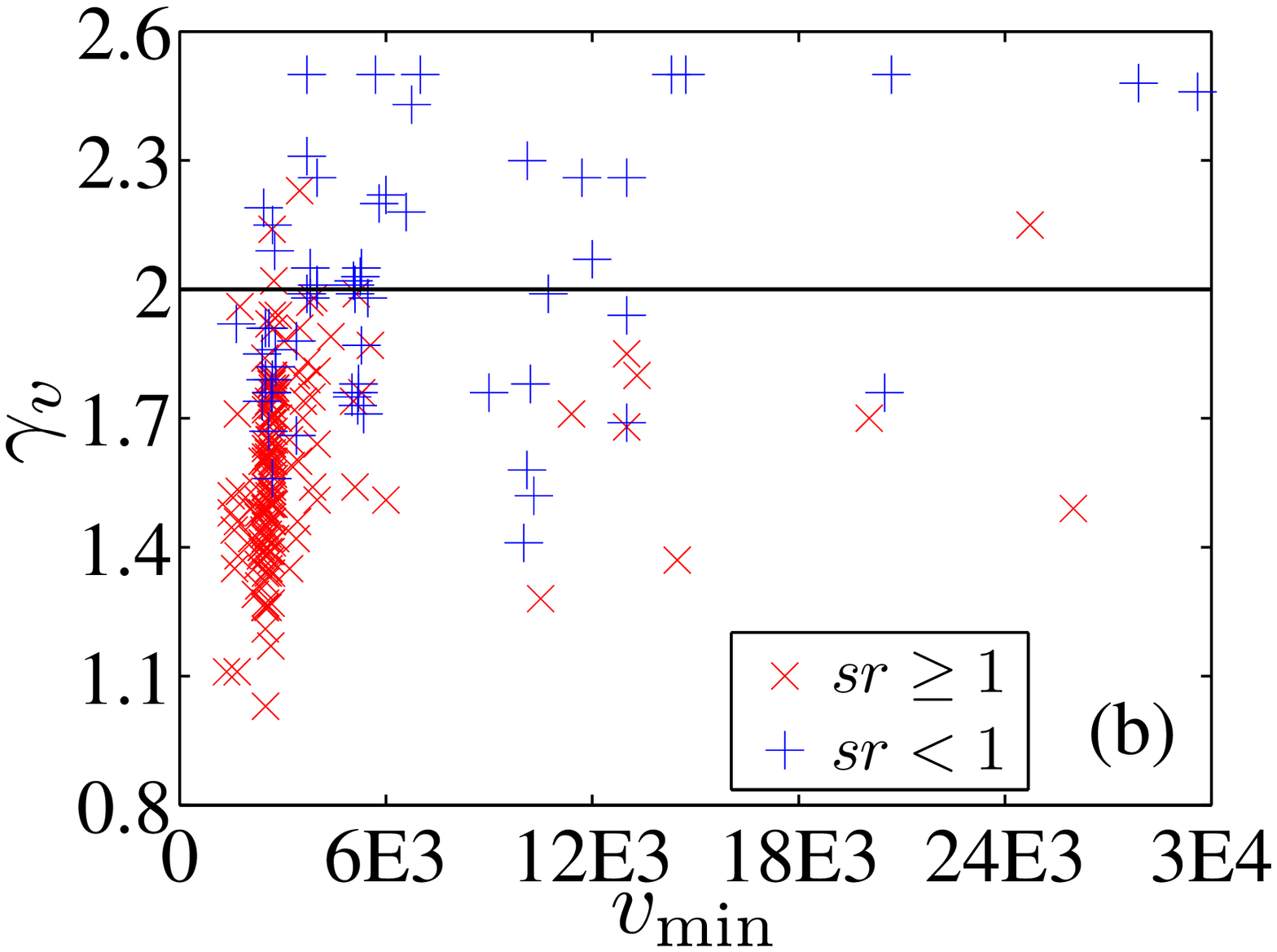}
\includegraphics[width=5cm]{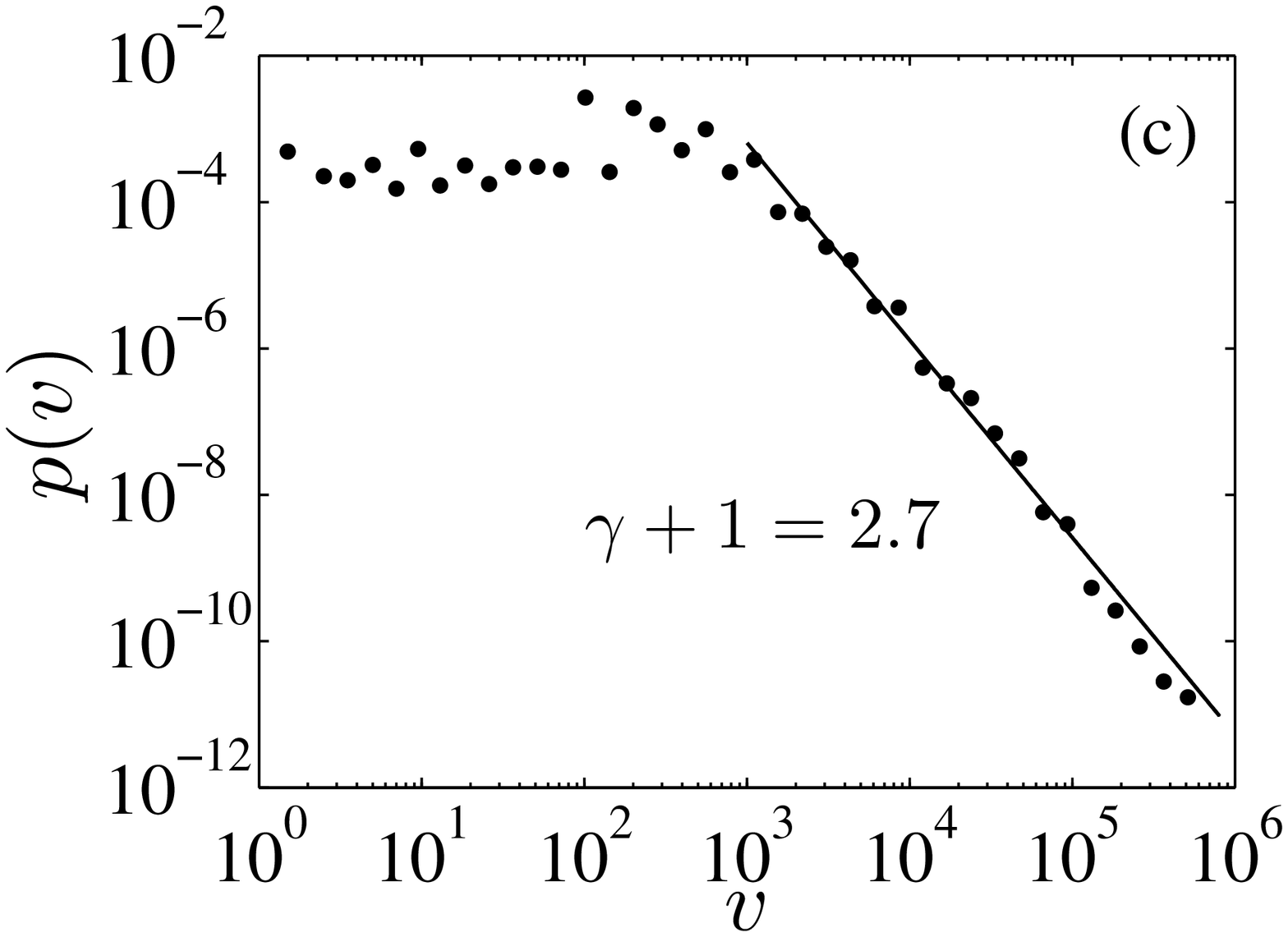}
\caption{\label{Fig:CDF:TradeSize} (Color online) Distribution of trade sizes $v$. (a) Cumulative probability distribution of trade sizes for three typical dates (January-03-2003, April-10-2003, and October-20-2003). The solid lines are the best fits to the power-law tails. (b) Scatter plot of the pairs $(v_{\min}, \gamma_v)$ for different trading days. Each point corresponds to the low boundary of power law $v_{\min}$ and the power law exponent $\gamma_v$. The marker $\times$ means the trading day on which the scaling range for power-law tail is not less than one order of magnitude and the marker $+$ represents the trading day on which the scaling range is less than one order of magnitude. (c) Empirical probability distribution of trade sizes $v$ over entire investigated period. The solid line is the best fit to a power-law distribution.}
\end{figure}

Fig.~\ref{Fig:CDF:TradeSize}(a) illustrates the cumulative probability distribution of trade sizes $v$ for three typical trading days (January-03-2003, April-10-2003, and October-20-2003). One can see that there is an asymptotical power-law decay when $v$ is larger than $v_{\min}$, such that
\begin{equation}
 C(v) \sim v^{-\gamma_v},~~~~{\rm{if}}~v \ge v_{\min}.
 \label{Eq:PL:TradeSize}
\end{equation}
In order to confirm our observations of power-law trade size distributions, a strategy is adopted as a standard procedure to check whether a given sample is drawn from a specific distribution, which is to fit the empirical data with maximum likelihood estimation and test the goodness of fits with the Kolmogorov-Smirnov (KS) statistic. Recently, Clauset, Shalizi and Newman (CSN) have proposed a power-law fitting approach for estimating the power-law exponent of empirical data as well as the low bound of the power-law behavior \cite{Clauset-Shalizi-Newman-2009-SIAMR}. The approach combines maximum-likelihood fitting methods with goodness-of-fit tests based on the Kolmogorov-Smirnov statistic and likelihood ratios.

We proceed the fitting procedure on the trade size samples by applying the above-mentioned powerful power-law calibrating technique for all the trading days. Calibrating results of three typical days are chosen as examples to illustrate the power-law tail distributions. As illustrated in Fig.~\ref{Fig:CDF:TradeSize}(a), the open markers are the empirical distributions and the solid lines are the best fits to the power-law distributions, which gives $\gamma_v = 1.76$ ($v_{\min} = 2749$) for January-03-2003, $\gamma_v = 1.27$ ($v_{\min} = 2500$) for April-10-2003, and $\gamma_v = 1.31$ ($v_{\min} = 2300$) for October-20-2003, respectively. The obtained calibrating parameters $v_{\min}$ and $\gamma_v$ are then substituted into the KS test procedure to test the goodness of fitting. The null hypothesis $H_0$ for our KS test is that the data $(v \ge v_{\min})$ are drawn from a power-law distribution. We find that the null hypothesis $H_0$ cannot be rejected at the significant level of 0.01 for all the days. Fig.~\ref{Fig:CDF:TradeSize} (c) depicts the bivariate distribution of pairs ($v_{\min}, \gamma_v$). Each node is associated with the low bound of power-law tail $v_{\min}$ and the estimated power-law exponent $\gamma_v$. We utilize the notation $sr$ to represent the scaling range of the power-law tail and distinguish $sr \ge 1$ and $sr < 1$ by marker $\times$ and $+$. $sr \ge 1$ means the scaling range is not less than one order of magnitude, and $sr < 1$ represents the opposite. The percentage of trading days on which the scaling range of the trade sizes spans not less than one order of magnitude is around 73.42\% in the samples. Concerning the trading days with the scaling ranges no less than one order of magnitude, there are 97.70\% days whose power-law exponents are well within the L{\'e}vy-stable regime. In addition, we also study the probability distribution of the trade sizes over the entire time period investigated, as illustrated in Fig.~\ref{Fig:CDF:TradeSize} (c). A nice power-law tail distribution is observed and maximum likelihood estimate also gives the tail exponent $\gamma = 1.7$, which is well inside L{\'e}vy stable regime.

\section{Stock trading networks}
\label{Sec:Network}

\subsection{Construction}

\begin{figure}[b!]
 \centering
 \includegraphics[width=6cm]{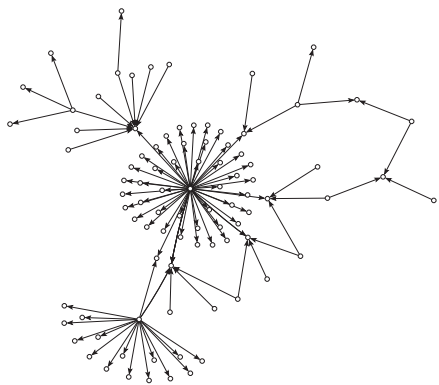}
 \caption{\label{Fig:TN} Part of the directed stock trading network constructed from the limit order book data on October 9, 2003. The nodes represent investors in the market and the directed edges stand for transactions.}
\end{figure}

In order to construct the stock trading networks, we have to reconstruct the limit order book based on the trading rules and to extract the detailed information of each transaction. A transaction is triggered by an incoming market order matched with the limit orders waiting on the opposite order book and accomplished by transferring shares from seller to buyer and cashes from buyer to seller, as sketchily illustrated in Fig.~\ref{Fig:TradeSize}. It provides an opportunity to trace the order execution procedure from a complex network perspective. Hence, we can mathematically write the trade sizes, defined in Sec.~\ref{Sec:Transaction}, between pairwise traders into a $N \times N$ matrix $V_t$ on trading day $t$. Its entry $v_{ij,t}$ corresponds to the trade size, which represents the number of shares flowing from investor $i$ to investor $j$. We only take into account directed networks here, therefore we have $v_{ij,t} \ne v_{ji,t}$. The adjacent matrix is defined as $A_t = [a_{ij,t}]$, where $a_{ij,t} = 1$ if $v_{ij,t} > 0$ and $a_{ij,t} = 0$ if $v_{ij,t} = 0$. A schematic diagram is illustrated in Fig.~\ref{Fig:TN}, which is a part of the trading network constructed from the limit order book data on October 9, 2003.

\subsection{Evolution of network size}

\begin{figure}[htb]
\centering
\includegraphics[width=8cm]{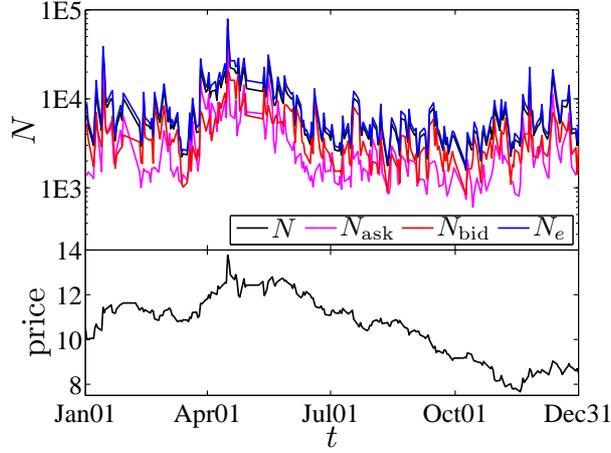}
\caption{\label{Fig:NetworkSize} (Color online) Plots of the number of traders $N$, ask-traders $N_{a}$, bid-traders $N_{b}$, and trades $N_e$ in the up panel. For comparison, the stock prices are also plotted in the low panel.}
\end{figure}

The network size varies with respect to the trading days $t$, for no two arbitrary days share the same trading activities. We utilize four quantities to describe the network size: the number $N$ of traders, the number $N_{\rm{ask}}$ of traders who submitted ask orders, the number $N_{\rm{bid}}$ of traders who submitted bid orders, the number $N_e$ of trades (or edges in the trading network). We have an inequality $N \leqslant N_{\rm{ask}} + N_{\rm{bid}}$, since it is possible that some traders submit both buy and sell orders on the same day. We plot the number of traders $N$, ask-traders $N_{\rm{ask}}$, bid-traders $N_{\rm{bid}}$, and trading edges $N_e$ with respect to the trading day $t$ in the up panel of Fig.~\ref{Fig:NetworkSize}. From the observation that the $N$ and $N_e$ curves almost are overlapping together, one can infer that $N_e$ is a little larger than $N$, which indicates that the constructed network resembles a tree-like network. This leads to the fact that the clustering coefficient of trading networks is very low. We also present the prices of our investigated stock in the corresponding time interval and find that $N$, $N_{\rm{ask}}$, $N_{\rm{bid}}$, and $N_e$ are all synchronous with the stock prices. The corresponding correlation coefficients are 0.44 for price and $N$, 0.36 for price and $N_{\rm{ask}}$, 0.45 for price and $N_{\rm{bid}}$, and 0.44 for price and $N_e$, respectively.

\subsection{Largest component}

\begin{figure}[htb]
\centering
\includegraphics[width=8cm]{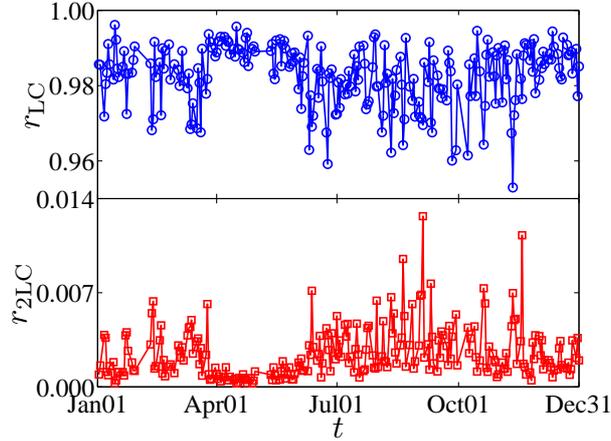}
\caption{\label{Fig:NetworkLC} (Color online) Plots of the relative sizes of the largest component $r_{\rm{LC}}$ (up panel) and the second largest component $r_{\rm{2LC}}$ (low panel).}
\end{figure}

Many real world networks are fragmented into isolated subnetworks or ``components''. A component is a part of a network, in which two arbitrary nodes are connected by at least one path, while there are no edges connecting two different components. Among these components, the largest component which encompasses the largest number of nodes is considered as the most important component in the network. The evolution of the relative size $r_{\rm{LC}}$ of the largest component that is defined as the ratio of the number of nodes in the largest component to the total number of nodes in a network is illustrated in the up panel of Fig.~\ref{Fig:NetworkLC}. We also show as the relative size $r_{\rm{2LC}}$ of the second largest component for comparison. One finds that the minimum $r_{\rm{LC}}$ is greater than 95\% in our investigated samples. In contrast, the second largest component only contains a very small fraction of the nodes in the system. Our analysis indicates that most of the traders are connected by a giant component. It is even interesting to investigate the correlation relationship between $r_{\rm{LC}}$ (also $r_{\rm{2LC}}$) and financial variables, including volatility and total trading volume. We employ the price range defined as the difference of maximum price and minimum price (in logs) as the measure of volatility for each trading day. The correlation coefficients are 0.42 for $r_{\rm{LC}}$ and volatility, -0.30 for $r_{\rm{2LC}}$ and volatility, 0.42 for $r_{\rm{LC}}$ and volume, -0.29 for $r_{\rm{2LC}}$ and volume, respectively.

\section{Network degrees}
\label{Sec:Degree}

Degree is regarded as one of most important quantities in complex networks, which describes how many neighbors each node has. Here, in our trading networks, degree can be interpreted as the number of trading counterparties of each investor. Because our constructed trading network is a directed network, we define three types of degree to illustrate the trade relationship in stock market, including the ask-degree (out degree) $k_{\rm{ask}} = \sum_i A_{ij}$, the bid-degree (in degree) $k_{\rm{bid}} = \sum_j A_{ij}$, and the total degree $k = k_{\rm{ask}} + k_{\rm{bid}}$ for each trader. We note that $\sum k_{\rm{ask}} = \sum k_{\rm{bid}}$ for all traders due to the buy-sell parity.

\subsection{Evolution of average degrees}

\begin{figure}[htb]
\centering
\includegraphics[width=8cm]{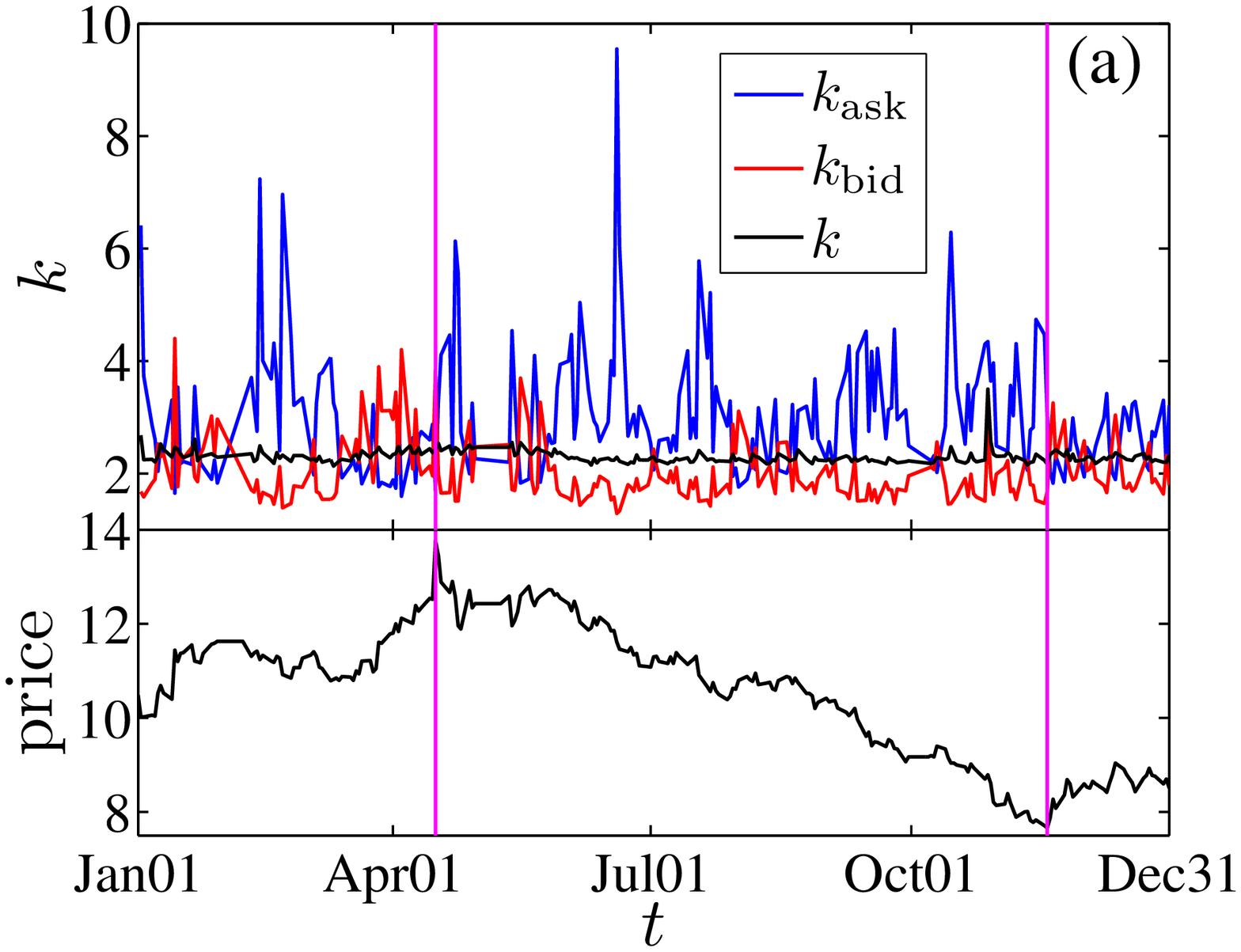}
\includegraphics[width=8cm]{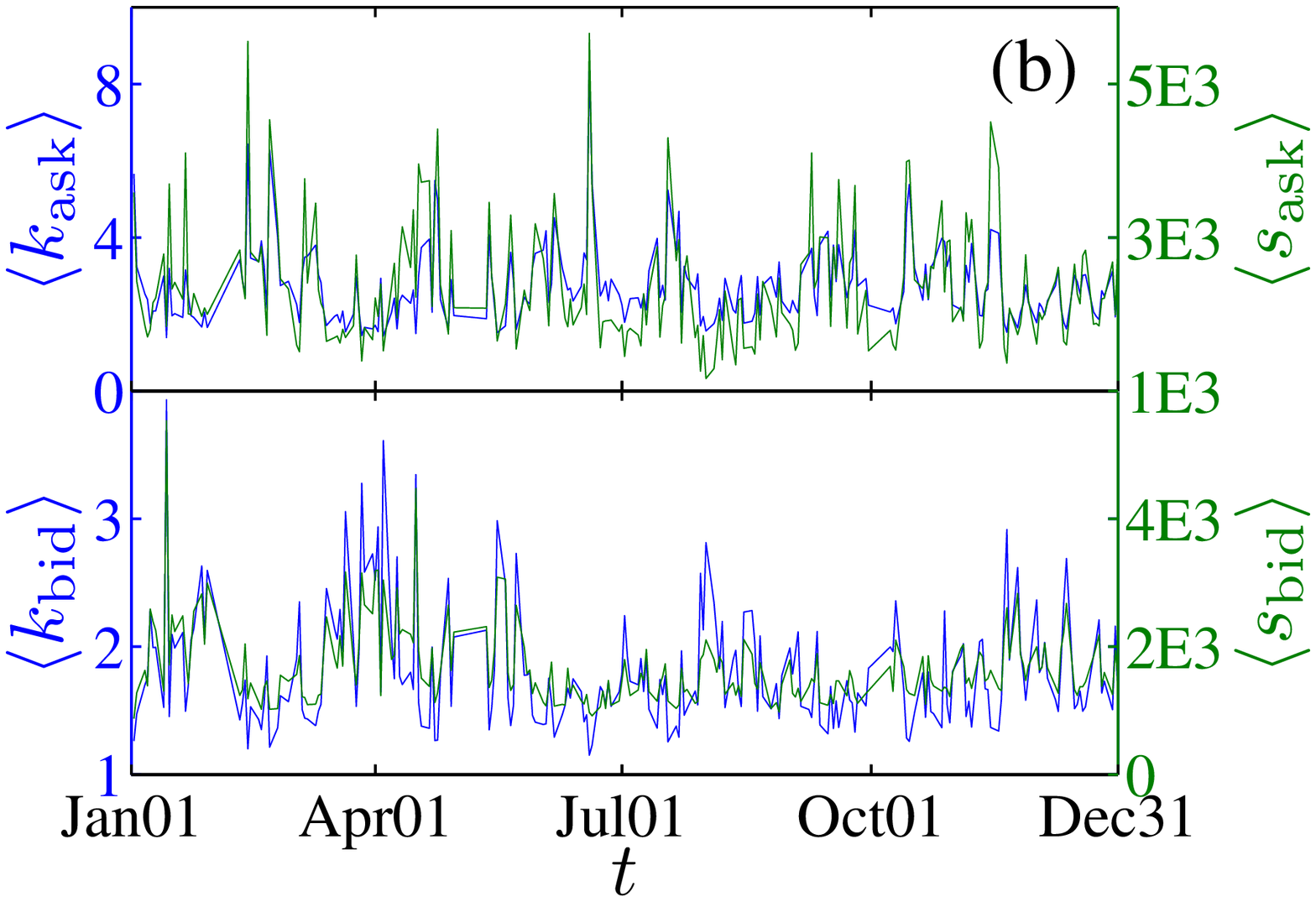}
\caption{\label{Fig:NetworkMK} (Color online) Analysis of average degrees. (a) Evolution of the average bid-degree $\langle k_{\rm{ask}} \rangle$, ask-degree $\langle k_{\rm{bid}} \rangle$, and total degree $\langle k \rangle$ with respect to the trading days. For illustration of bearish and bullish period, separated by two vertical lines, the stock price is presented in the low panel. (b) Pairs of [$\langle k_{\rm{ask}} \rangle$ (left y-axes), $\langle s_{\rm{ask}} \rangle$ (right y-axes)] and [$\langle k_{\rm{bid}} \rangle$ (left y-axes), $\langle s_{\rm{bid}} \rangle$ (right y-axes)] evolving with the trading days.}
\end{figure}

Fig.~\ref{Fig:NetworkMK} (a) illustrates the evolution of the daily average values of $\langle k_{\rm{ask}} \rangle$, $\langle k_{\rm{bid}} \rangle$, and $\langle k \rangle$. Comparing with the stock prices, one can see that $\langle k_{\rm{ask}} \rangle > \langle k_{\rm{bid}} \rangle$ in the bearish range and $\langle k_{\rm{bid}} \rangle > \langle k_{\rm{ask}} \rangle$ in the bullish ranges. We also observe that there are many sharp spikes for $\langle k_{\rm{bid}} \rangle$ and especially for $\langle k_{\rm{ask}} \rangle$. In order to give an interpretation of this phenomenon, we plot the daily average degrees and the daily average order sizes $\langle s \rangle$ on the same coordinates. As illustrated in Fig.~\ref{Fig:NetworkMK} (b), pairs of [$\langle k_{\rm{ask}} \rangle$ (left y-axes), $\langle s_{\rm{ask}} \rangle$ (right y-axes)] are sketched in the upper panel and pairs of [$\langle k_{\rm{bid}} \rangle$ (left y-axes), $\langle s_{\rm{bid}} \rangle$ (right y-axes)] are presented in the lower panel. These sharp spikes are a consequence of the appearance of relatively large values of average order sizes. However, the average values of $\langle k \rangle$ do not have as much fluctuations as $\langle k_{\rm{ask}} \rangle$ and $\langle k_{\rm{bid}} \rangle $, which has a less fluctuated value close to $2.29 \pm 0.12$ (mean $\pm$ std). This is due to the facts that the average degrees can also be calculated from $\langle k \rangle = 2N_e/N$ and that the two curves of $N_e$ and $N$ in Fig.~\ref{Fig:NetworkSize} almost superpose.

\subsection{Probability distribution}

\begin{figure}[htb]
\centering
\includegraphics[width=8cm]{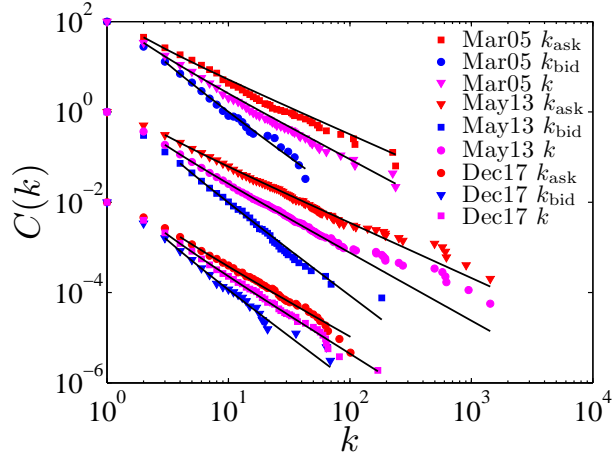}
\caption{\label{Fig:CDF:NetworkDegree:A} (Color online) Degree distribution distribution for three typical days (Mar-05-2003, May-13-2003, and Dec-17-2003). The filled markers are the empirical distribution and the solid lines are the best power-law fits to data according to the CSN approach. The data for Mar-05-2003 and Dec-17-2003 are translated by a factor of 100 and 0.01 for better visibility. }
\end{figure}

After understanding the evolution behavior of the average degrees, we further investigate the probability distribution of the three kinds of degrees for each trading day. Fig.~\ref{Fig:CDF:NetworkDegree:A} illustrates the cumulative distribution of $k_{\rm{ask}}$, $k_{\rm{bid}}$, and $k$ for three randomly chosen days (Mar-05-2003, May-13-2003, and Dec-17-2003). We can find that the cumulative probability distribution function $C(k)$ asymptotically decays as a power law with respect to $k$,
\begin{equation}
 C(k_{\rm{ask}}) \sim k_{\rm{ask}}^{-\gamma_{\rm{ask}}},~~C(k_{\rm{ask}}) \sim k_{\rm{bid}}^{-\gamma_{\rm{bid}}},~~C(k) \sim k^{-\gamma}.
 \label{Eq:PL:TradeSize:ask}
\end{equation}
We adopt the same strategy utilized in Sec.~\ref{Sec:Transaction} to confirm our observations of power-law degree distributions. First, we calibrate the empirical data by the CSN approach \cite{Clauset-Shalizi-Newman-2009-SIAMR}. Second, we perform KS test for judging the goodness-of-fit of the fitting distribution compared with the empirical distribution. The filled markers are the empirical distributions and the solid lines are the best fits to the power-law function form, which gives $\gamma_{\rm{ask}} = 1.32$ ($k_{\rm{ask},~\min} = 3$), $\gamma_{\rm{bid}} = 1.93$ ($k_{\rm{bid},~\min} = 3$), and $\gamma = 1.56$ ($k_{\min} = 3$) for Mar-05-2003, $\gamma_{\rm{ask}} = 1.21$ ($k_{\rm{ask},~\min} = 3$), $\gamma_{\rm{bid}} = 1.05$ ($k_{\rm{bid},~\min} = 5$), and $\gamma = 1.49$ ($k_{\min} = 3$) for May-13-2003, and $\gamma_{\rm{ask}} = 1.51$ ($k_{\rm{ask},~\min} = 4$), $\gamma_{\rm{bid}} = 1.96$ ($k_{\rm{bid},~\min} = 3$), and $\gamma = 1.7$ ($k_{\min} = 4$) for Dec-17-2003, respectively.

\begin{figure}[htb]
 \centering
 \includegraphics[width=8cm]{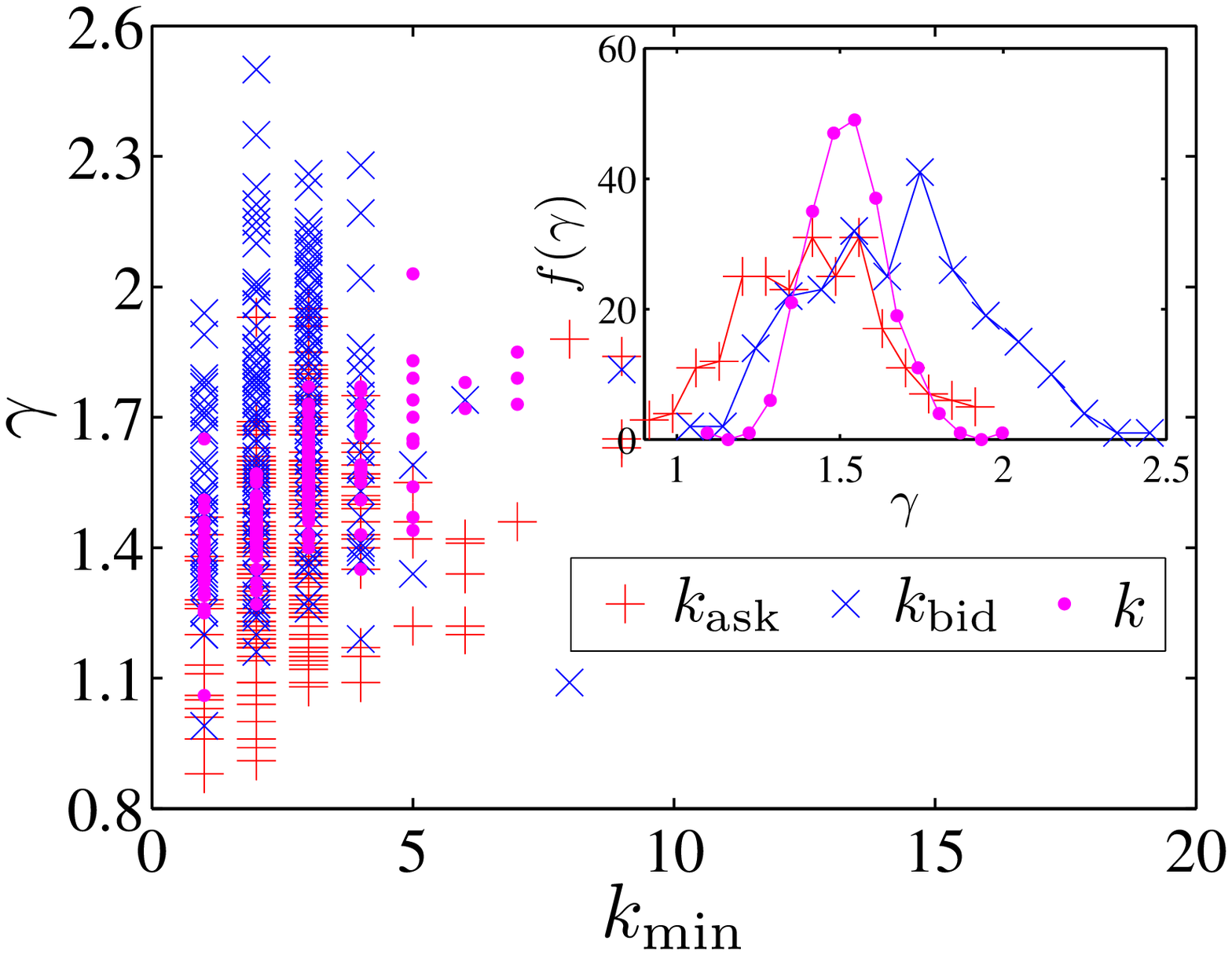}
 \caption{\label{Fig:CDF:NetworkDegree:B} (Color online) Bivariate distribution of pairs $(k_{\min}, \gamma)$ for different trading days. Each point corresponds to the low boundary of power law $k_{\min}$ and the power law exponent $\gamma$. The inset shows the empirical frequency of $\gamma$.}
\end{figure}

The KS test is implemented on the calibrating results to check the goodness-of-fit. Our null hypothesis $H_0$ is the degrees can be well modeled by a power-law distribution. At a significant level of 0.01, we find that the null hypothesis $H_0$ cannot be rejected for a fraction 100\% (99.6\%, 98.7\%, respectively) of the trading days (237 days in total) for ask-degree $k_{\rm{ask}}$ (bid-degree $k_{\rm{bid}}$, total degree $k$). Fig.~\ref{Fig:CDF:NetworkDegree:B} depicts the bivariate distribution of pairs ($k_{\min}, \gamma$) for the samples which pass the KS test. Each node is associated with the low bound of power-law tail $k_{\min}$ and the estimated power-law exponent $\gamma$. The inset shows the frequency of power-law exponents. The average values of the power-law exponents for the whole samples are $\gamma_{\rm{ask}} = 1.42 \pm 0.22$ (mean $\pm$ std), $\gamma_{\rm{bid}} = 1.69 \pm 0.26$, and $\gamma = 1.52 \pm 0.12$, which are well within the L{\'e}vy stable regime. We notice that the degree distribution of the directed Austrian trading network also falls in the L{\'e}vy stable regime \cite{Kyriakopoulos-Thurner-Puhr-Schmitz-2009-EPJB}.

\subsection{Disassortativity}

The average degree of nearest neighbor $k_{nn}(k)$ for vertices with degree $k$ is usually adopted to quantitatively characterize the architecture of the investigated networks. If $k_{nn}(k)$ has an ascending (respectively, descending) form with respect to $k$, one can argue that the network is assortative (disassortative). The average degree of nearest neighbor $k_{nn,i}$ for node $i$ is defined by
\begin{equation}
 k_{nn,i} = \frac{\sum k_j}{k}
 \label{Eq:dsn:knni}
\end{equation}
where $k$ is the degree of node $i$ and $k_j$ stands for the degree of the $j$-th neighbor of node $i$. Then the values of $k_{nn}$ can be obtained through averaging over $k_{nn,i}$ of all the nodes with the same degree $k$.

\begin{figure}[htb]
\centering
\includegraphics[width=8cm]{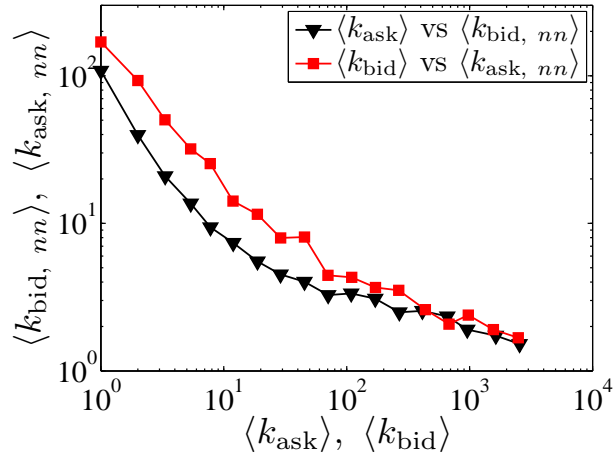}
\caption{\label{Fig:NetworkKNN} (Color online) Plots of $\langle k_{{\rm{bid}},~nn} \rangle$ (respectively, $\langle k_{{\rm{ask}},~nn} \rangle$) with respect to $k_{\rm{ask}}$ ($k_{\rm{bid}}$). The decreasing curves indicate that the stock trading networks have disassortative architectures.}
\end{figure}

In order to gain better statistics, we regard all the nodes from different networks as an ensemble set, which is partitioned into 18 non-overlapping groups. In the partitioning procedure, we equally divide the interval $[\ln[\min(k)], \ln[\max(k)]]$ into 18 groups and assign each node into the corresponding group if the degree of the node locate in the interval of the group. In each group, the average value $\langle k \rangle$ of $k$ and the associated average value $\langle k_{nn} \rangle$ of $k_{nn,i}$ are determined. In our case, the neighbors of asking (bidding) nodes are bidding (asking) nodes. The relationship between $\langle k_{nn} \rangle$ and $\langle k \rangle$ is plotted in Fig.~\ref{Fig:NetworkKNN} for asking nodes and bidding nodes. Both curves share the same decreasing form, which implies the disassortative configuration in the trading networks.

\subsection{Correlation between order sizes and degrees}

\begin{figure}[htb]
\centering
\includegraphics[width=8cm]{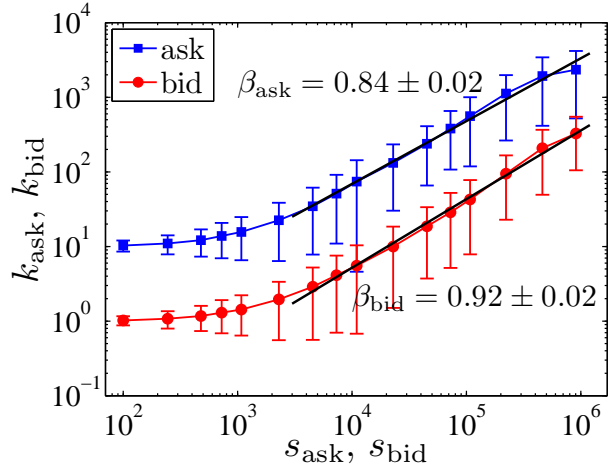}
\caption{\label{Fig:NetworkKOS} (Color online) Plots of the correlation between order sizes $s$ and degrees $k$ for buy and sell orders. The data for sell orders are shifted by a factor of 10 for better visibility. The solid lines are the best power-law fits to the data.}
\end{figure}

The trading host is a queuing system. The limit orders are sorted according to the principle of price-time priority in the limit order book. A transaction occurs when an incoming order matches with the limit order waiting on the opposed side. Normally, the market orders with large sizes will eat up a relatively large number of limit orders at the opposite side and the limit orders with lager size could consume a relatively large number of market orders as well. This may lead to the correlation between the network degrees and the order sizes. In order to verify this conjecture, we extract the information about the total order size $s$ and degree $k$ for each stock trader on each trading day. To gain better statistics, all the records of different trading days are regarded as an ensemble set. We logarithmically equally partition the interval $[\min(s), \max(s)]$ into 24 groups and assign each trader into the corresponding group if the order size locate in the interval of the group. For each group, the mean and standard deviation of order sizes and degrees are determined. However, we only get 16 data points for other 8 groups are empty sets. Fig.~\ref{Fig:NetworkKOS} illustrates the errorbar plots of average degrees with respect to average order sizes for 16 groups in double logarithmic coordinates. The order size $s$ exhibits an asymptotical power-law behavior with respect to $k$ when $s > 3000$,
\begin{equation}
 k \sim s^{\beta}.
\label{E:pl:ks}
\end{equation}
The least-squares fits to the data provide an estimation of the power-law exponents, such that $\beta_{\rm{bid}} = 0.92 \pm 0.02$ for buy orders and $\beta_{\rm{ask}} = 0.84 \pm 0.02$ for sell orders. This analysis demonstrates the positive correlation between order sizes and degrees, which inspires us to regard the order size as a fitness variable in fitness model and to give an explanation of the power-law degree distributions in the trading networks \cite{Caldarelli-Capocci-DeLosRios-Munoz-2002-PRL,Servedio-Caldarelli-Butta-2004-PRE,Garlaschelli-Loffredo-2004-PRL}.

\section{Fitness model}
\label{Sec:Model}

Variety of empirical complex networks extracted from different fields exhibit power-law degree distributions, and two fundamental mechanisms are proposed for the understanding of the ubiquitously emerging scaling-free behaviors. One is ``rich-get-richer'', and the other is ``good-get-richer''. The concept of ``rich-get-richer'', also known as {\em{preferential attachment}}, is well captured by the Barab{\'a}si-Albert model, in which networks proliferate at a constant rate and new nodes are added to existed nodes with a probability $p(k)$. The connected probability $p(k)$ of each node is linearly increasing to its degree, which indicates the nodes with high degrees will have high priority to acquire links from newly arriving nodes \cite{Barabasi-Albert-1999-Science}. The model of ``good-get-richer'' is introduced to interpret some empirical networks with scale-free behavior stemming from the vertex intrinsic fitness \cite{Caldarelli-Capocci-DeLosRios-Munoz-2002-PRL}. Garlaschelli and Loffredo have utilized this model by assigning the gross domestic product (GDP) as the vertex fitness to each node, which is able to reproduce the topological properties of world trade webs \cite{Garlaschelli-Loffredo-2004-PRL}.

From the observation of the power-law correlation between order size $s$ and degree $k$, we conjecture that the appearance of power-law degree distributions results from the order sizes. To verify our conjecture, the fitness model is adopted to simulate the stock trading networks. We define a set of nodes with fitnesses, in which the nodes are the investors and the fitnesses are their order sizes $s$ from the original order flow data. The network-building algorithm is proposed as follows. Two nodes $i$ and $j$ are randomly chosen from the set and a transaction occurs between the two nodes whose trade size is determined by the smaller order, $v = \min(s_i, s_j)$. Then we subtract the trade size from the two original order sizes, $s_i = s_i - v$ and $s_j = s_j - v$. If the remaining order size equals to zero, the corresponding node is removed from the set. After all investors are removed, the process ceases.

\begin{figure}[htb]
\centering
\includegraphics[width=7cm]{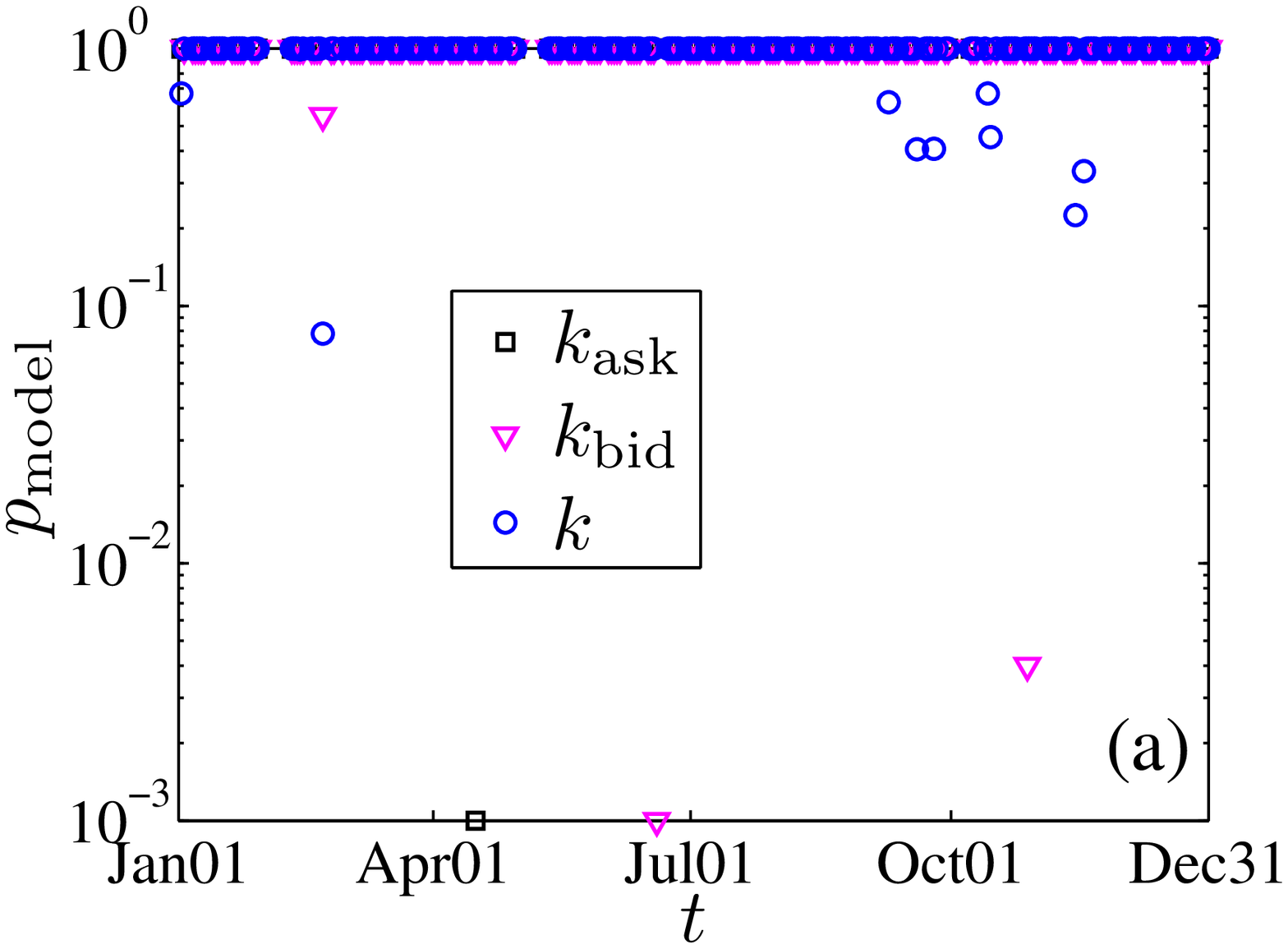}
\includegraphics[width=7cm]{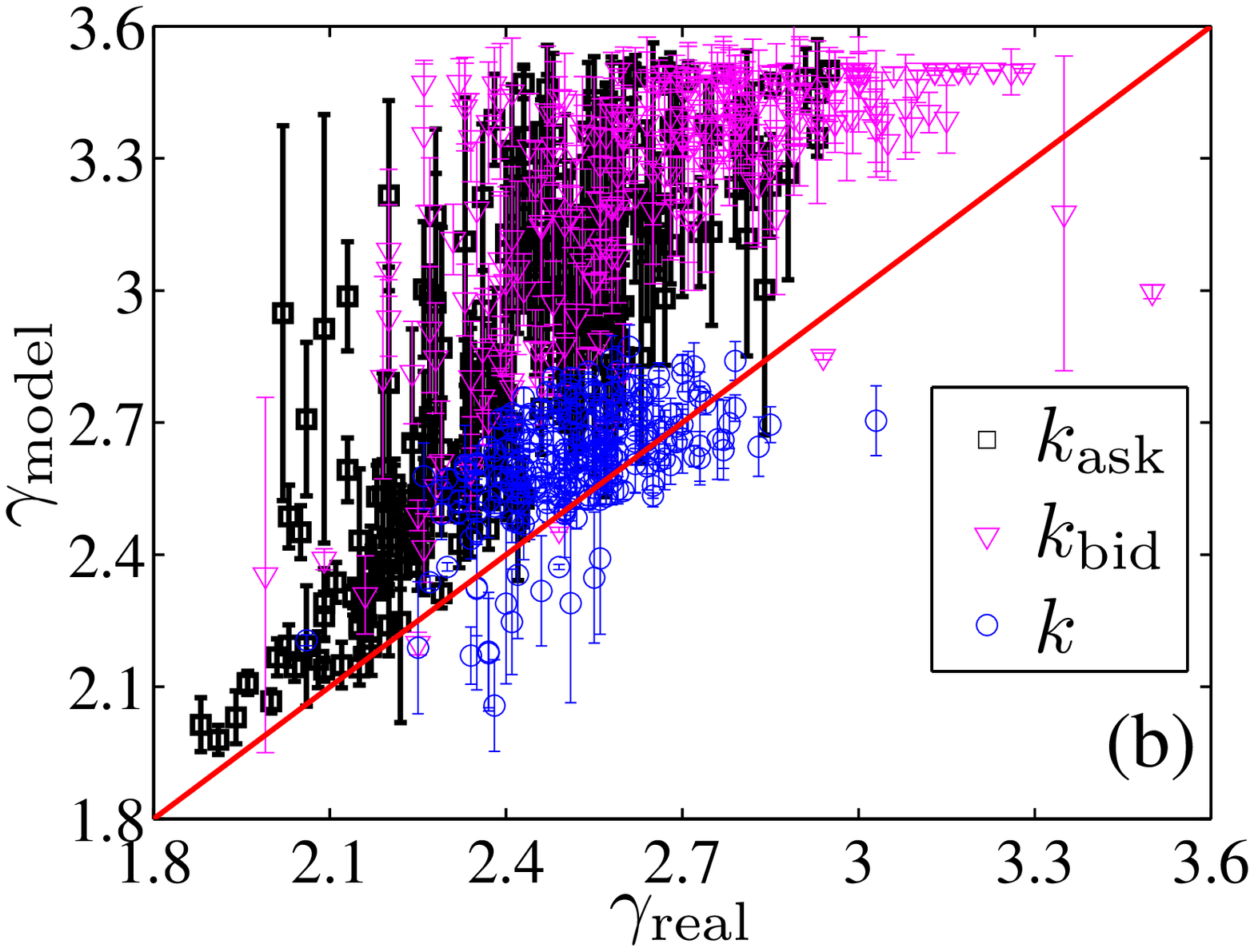}
\caption{\label{Fig:NetworkModel} Results of the fitness model. (a) Plots of the percentage $p_{\rm{model}}$ with respect to the trading days $t$. (b) Errorbar plots of $\gamma_{\rm{real}}$ as a function of $\gamma_{\rm{model}}$. The solid line represents $\gamma_{\rm{model}} = \gamma_{\rm{real}}$.}
\end{figure}

For each trading day, we obtain 1000 synthetic networks generated from the network-building algorithm. We adopt the same strategy utilized in Section \ref{Sec:Transaction} to check whether the degrees of each synthetic network are drawn from a power-law distribution. We define $p_{\rm{model}}$ as the percentage of the synthetic networks whose degree distributions pass the KS test at a significant level of 0.01 for a given trading day. As illustrated in Fig.~\ref{Fig:NetworkModel} (a), the percentage $p_{\rm{model}}$ is plotted with respect to the trading days $t$ for ask-degree $k_{\rm{ask}}$, bid-degree $k_{\rm{bid}}$, and degree $k$, respectively. We find a fraction 99.6\% (respectively, 97.9\%, 95.4\%) of the trading days (237 days in total) on which the percentage $p_{\rm{model}}$ is larger than 99\% for ask-degree $k_{\rm{ask}}$ (bid-degree $k_{\rm{bid}}$, total degree $k$). Fig.~\ref{Fig:NetworkModel} (b) presents the scatter plots of $\gamma_{\rm{real}}$ with respect to $\gamma_{\rm{model}}$. One can observe that most of the points locate above the solid line $\gamma_{\rm{model}} = \gamma_{\rm{real}}$, indicating that the power-law exponents obtained from fitness model are lager than that of real market. The deviations between $\gamma_{\rm{real}}$ and $\gamma_{\rm{model}}$ reflect the fact that the considerations of the sequence and the ask or bid prices are also important in the matching mechanism.

\section{Concluding remarks}
\label{Sec:Conclusion}

In this paper, we have performed an investigation of the statistical properties of stock trading networks based on the order flow data of a high-liquidity stock listed on Shenzhen Stock Exchange (Shenzhen Development Bank) during the whole year of 2003.

For each trading day, we have reconstructed the limit order book and extracted the detailed information of each executed order (including the order matching, trade price, and the corresponding trade size). The trade sizes are found to exhibit power-law distributions and most of the estimated power-law exponents are well inside the L{\'e}vy regime.

Based on the detailed trading information, one can establish stock trading networks, in which the investors represent nodes and each transaction is translated as an directed edge drawn from a seller to a buyer with the trade size as its weight. Each network contains a largest component, which contains more than 95\% of investors. Furthermore, the trading networks surprisingly have power-law degree distributions and disassortative architectures. The fitness model is employed to interpret the consequence of power-law distributions, which are motivated by the power-law correlation between order sizes and degrees.

Trading networks have provided an opportunity to trace the order execution process in financial markets from the perspective of complex networks. Future works for the application of trading networks are necessary to be carried out. For instance, one can try to investigate the relations between network variables and traditional financial variables, between local network structures and abnormal trading behaviors, between network eigenvalues and abnormal traders, and so on. We believe that trading network analysis is a potentially useful tool for analyzing the microstructure of financial markets \cite{Lillo-Farmer-2004-SNDE,Farmer-Lillo-2004-QF,Farmer-Gillemot-Lillo-Mike-Sen-2004-QF,Gillemot-Farmer-Lillo-2006-QF,Mike-Farmer-2008-JEDC,LaSpada-Farmer-Lillo-2008-EPJB}.

\bigskip
{\textbf{Acknowledgments:}}
We are grateful to Fei Ren and Gao-Feng Gu for fruitful discussions. This work was partially supported by the ``Shu Guang'' project (Grant No. 2008SG29) and the ``Chen Guang'' project (Grant No. 2008CG37) sponsored by Shanghai Municipal Education Commission and Shanghai Education Development Foundation, and the Program for New Century Excellent Talents in University (Grant No. NCET-07-0288).

\bibliography{E:/Papers/Auxiliary/Bibliography}

\end{document}